# Emulating homoeostatic effects with metal-oxide memristors T-dependence

Thomas Abbey, Alexantrou Serb, Spyros Stathopoulos, Loukas Michalas, and Themis Prodromakis

**Memristor technologies have been rapidly maturing for the past decade to support the needs of emerging memory[1], artificial synapses[2,3], logic gates[4] and bio-signal processing[5] applications. So far, however, most concepts are developed by exploiting the tuneable resistive state of memristors with other physical characteristics being ignored. Here, we report on the thermal properties of metal-oxide memristors and demonstrate how these can be used to emulate a fundamental function of biological neurons: homoeostasis. We show that thermal control mechanisms, frequently dismissed for their generally slow and broad-brush granularity, may in fact be an appropriate approach to emulating similarly slow and broad-brush biological mechanisms due to their extreme simplicity of implementation. We further demonstrate that metal-oxide memristors can be utilised as thermometers and exhibit a programmable temperature sensitivity. This work paves the way towards future systems that employ the rich physical properties of memristors, beyond their electrical state-tuneability, to power a new generation of advanced electronics solutions.**

Neuromorphic computing and the strong push of AI technology is creating a strong need for efficient neural networks-on-chip. To that end, metal-oxide[6], phase-change[7] and spintronic[8] memristors have been undergoing intense developments as a replacement for the most common component of any neural network: the synapse. The driving force behind such synapse substitutes have been technology attributes such as stochasticity[9], high analogue resolution[10] and the promise of low power operation[11] and high integration density[12]. All such cases, exploit memristors' ability to change their resistance as a result of appropriate electrical biasing. However, the conduction mechanisms underlying metal-oxide, and potentially other memristive device technologies, also feature temperature dependence[13], which can be used to further increase the computational capability of memristive synapses. Specifically, the slow effect of temperature on conductance, which acts as the synaptic weight of a memristive device, appears well-suited for emulating homoeostatic effects in artificial neurons, i.e. allowing neurons to maintain high sensitivity to transient changes in input levels despite having limited operating output range and regardless of the running baseline level of input activity[14]. Homoeostasis in memristor-based artificial neural networks (ANNs) have been touched upon briefly in the past both theoretically[15] and practically[16], but in both cases the homoeostatic mechanism was implemented at the neural soma level, i.e. effectively excluding the memristive synapses from the process.

In this work, we characterise the thermal dependence of metal-oxide memristors alike the one depicted in Fig. 1a and use this property for demonstrating homoeostatic plasticity by relying on intrinsic thermal effects on the synapses. Specifically, we examine the temperature dependence of the resistive state of prototyped memristors and show that the derivative of measured resistance vs. temperature is a function of the current resistive state. Next, we show that the fractional (%) resistive switching does not show any clear signs of dependence on temperature, which implies that the learning rate, when normalised to a standard temperature, remains fixed regardless of which temperature plasticity is applied at. Finally, we demonstrate how the combined thermal effects can be used to introduce homoeostatic plasticity in an artificial neuron. Our results closely resemble biological neuron habituation where large step changes in combined input signal intensity lead to a combination of significant step responses followed by habituation to a slightly modulated baseline level. In other words, jumps in the input are registered as jumps in the output, but after the homoeostatic system settles the artificial neuron shows a small, but distinct change in baseline firing. This simultaneously allows a direct, low resolution read-out of the current input levels and primes the system for responding to future step changes in input signal with easily detectable transients, regardless of direction of input signal change.

**Static thermal effects in metal-oxide memristors**

To assess the effects of temperature on static resistance we applied a series of standardised readout voltage pulses at 0.2V, as temperature was set within a 300-360 K range in steps of 10K. The order of temperature application was scrambled as shown in Fig. 1b in order to reduce the probability of patterns being formed as an artifact resulting from monotonic, regular temperature ramping. Each temperature level was held for 1 hr, after which the measured resistance change per 6 minutes is well less than 2% of the total change since step-changing the applied temperature. During thermal cycling, we visited the 300K and 360K temperatures twice to check whether the cycling procedure affected the device under test (DUT) lastingly. This temperature range was chosen as it reflects the conditions endured in practical computing applications. The chosen read-out voltage lies far below the switching threshold, which for these particular specimens is defined as the voltage at which a 100 µs pulse induces a 2% resistivity change with respect to the baseline value at T=300 K. This protocol (Fig. S2a) was initially applied on pristine (un-electroformed) devices, as shown in Fig. 1b. A clear correlation between temperature and measured resistance is observed. We notice that despite allowing for sufficient saturation time the resistive state shows some small residual difference in resistance between the two visits to 300K and 360K, with the steady-state resistance dropping from ~1.8 MΩ to ~1.7 MΩ (difference of 5%). This figure provides an estimate of the repeatability of the measurement within a single device.

Repeating this protocol for multiple devices, each programmed at different levels of resistive state (1 device/resistive state), yielded the results shown in Fig. 1c. We note that the derivative of fractional resistive state change decreases in absolute value terms as the resistive state of the device drops. The difference in resistance modulation between devices programmed at lowest (8 kΩ) and highest (3MΩ, pristine cell) resistive states is approximately a factor of 6 (11% vs. 61%), indicating a clear state-dependent thermal sensitivity. Moreover, the corresponding thermionic emission plot, depicted in Fig. 1d, indicates that a consistent mechanism of electronic conduction applies. This reinforces our finding that one can tune the thermal sensitivity for a DUT by operating the device in a distinct resistive range that provides a key degree of freedom for the design of artificial neurons with thermal effect-based homoeostasis or tuneable memristor thermometers.

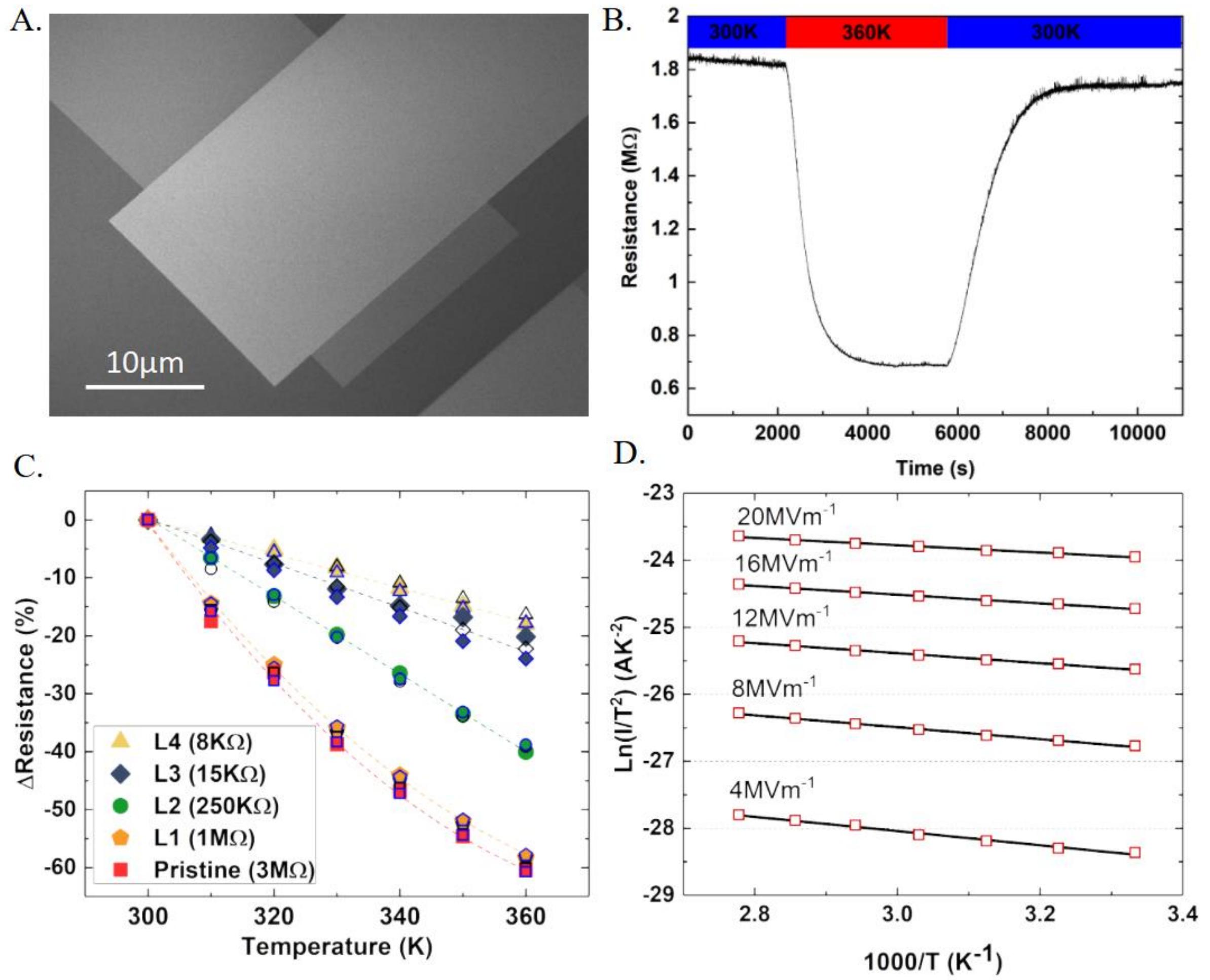

**Fig 1: Temperature dependence of a metal-oxide memristor under read-out conditions. a**, Scanning electron microscope micrograph of a prototype device. **b**, Transient response of the device's resistance under thermal cycling from 300 K to 360 K and back to 300 K. **c**, Temperature sensitivity of a metal-oxide memristor prototype programmed at distinct restive levels (L1-L4) indicating a state-dependent T sensitivity (derivatives of traces). **d**, Thermionic emission signature plot for a memristor prototype operating at L1, showing the dominance of interfacial conduction mechanism.

## Dynamic Thermal Effects in metal-oxide memristors

In the proposed system, homoeostatic plasticity is implemented by resourcing to thermal stimulation that directly affects the weights of the artificial synapses. As a result, we must ensure that the effective learning rate of the system does not depend on temperature, i.e. that the fractional change in resistive state caused by a plasticity event (long-term potentiation or depression[16,17]) is not a function of temperature. To that end we characterised the fractional change in resistive state after applying a standardised supra-threshold stimulation protocol at different temperatures. First, the DUT's resistance was assessed at the reference temperature of 300 K and this initial value was recorded. Then, the temperature was set to the desired test level and after waiting for 1h in order to stabilise the temperature in-silico. The devices were then stimulated with trains of 200 programming pulses at an amplitude of 1.5 V (well above their switching threshold) and duration of 100 µs. Then, a retention test was run using 200×100 µs pulses (see supplementary Fig. S2b). Device resistance was measured after each programming pulse and during the retention run. Following the application of this heat-stimulate-retention protocol, the temperature was returned to 300 K and the device's resistive state was reset to its initial 300 K reference level using negative voltage pulses before rerunning the test at a different temperature. This was done to ensure a consistent initial state for all runs at all temperatures and thus minimise any thermal impact.

The exemplar results shown in Fig. 2a were acquired from an L1 prototype where the initial memory state of the DUT was set to 1MΩ. With the exception of the 300 K run that was carried out first, it can be observed that the fractional change in resistance does not strongly depend on temperature: the fractional change in resistive state was within approximately 10% for all temperatures. Rather, the observed differences in fractional change are more strongly correlated to the order in which the temperature tests were run. The initial 300 K run was likely to be a result of burn-in; a phenomenon we have observed previously where the first round of stimulation has a more significant impact on the DUT. Next, we noted that the change in resistive state was volatile, that the recovery was incomplete by the end of the retention run and importantly, that the fractional recovery was also not strongly dependent on temperature. This experiment was repeated on the same prototype device for multiple pulse amplitudes ranging from 0.7V to 1.4V in steps of 0.1V. The resulting nullclines, summarised in Fig. 2b, confirm that there is a consistent, but small dependence of resistance modulation as a function of temperature, with the largest difference in observed fractional changes across temperature being measured at 1.4V bias where we gauge approximately 22% ΔR at 310 K and 27% at 360 K (the same case as highlighted in Fig. 2a).

Similar results for devices at different initial resistive states confirmed that the fractional change in resistance that is induced by a fixed pulse representing a plasticity event does not depend strongly on temperature regardless of current resistive state or whether the transition is volatile or non-volatile. We can therefore approximate that the learning rate is not affected by thermal effects and thus typical learning rules such as STDP, triplet[19] or other rules can be used normally so long as the system is not too sensitive to small possible changes in learning rate.

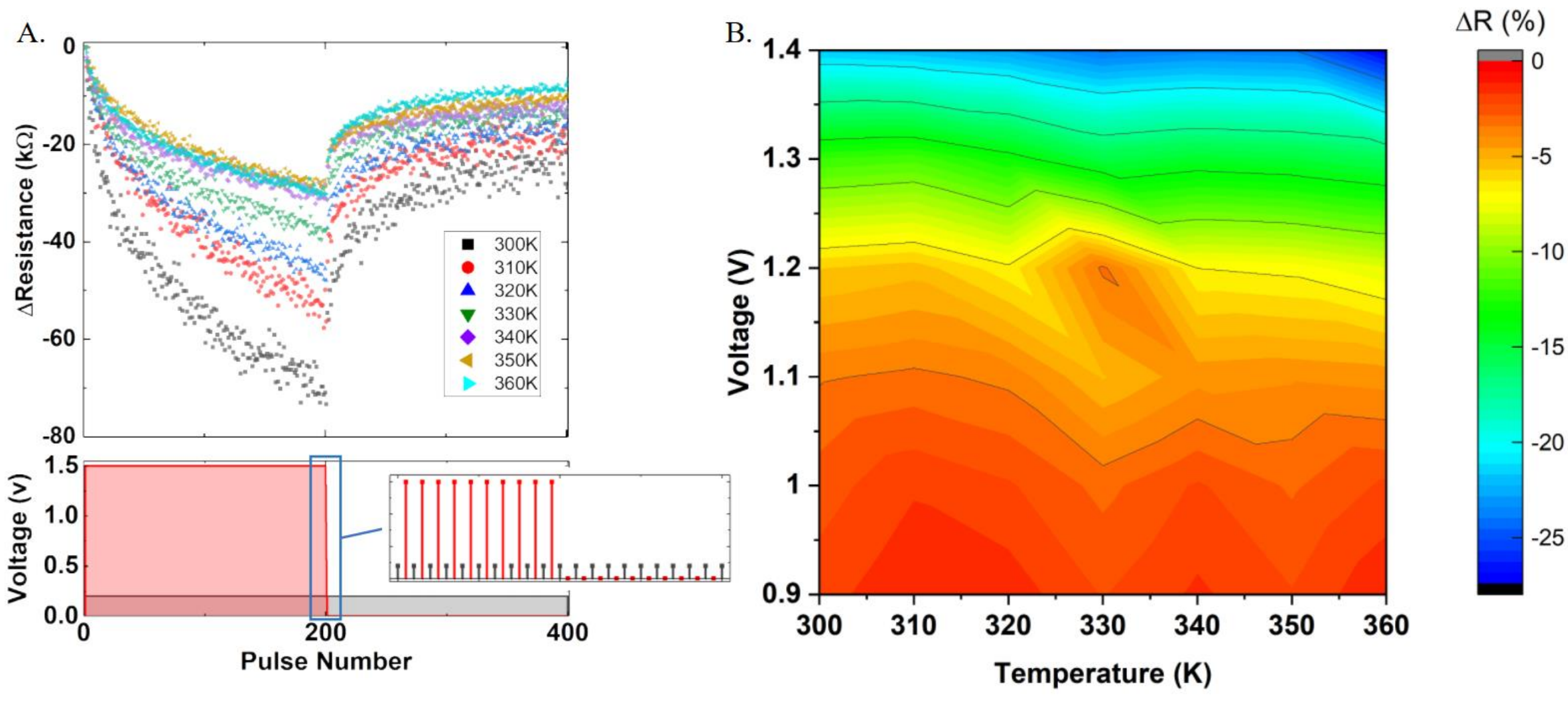

**Fig. 2: Temperature dependence of a metal-oxide memristor under switching conditions. a**, Normalised transient response of an L1 metal-oxide memristor's state to 200 {1.5 V, 100 μs} programming pulses followed by a 200-point retention with varying temperature. **b,** The effect of voltage and temperature on the switching behaviour of an L1 specimen, showing an increase in magnitude of the proportional switching behaviour.

## Emulating homoeostasis using thermal effects

The data in Fig. 1c, as well as additional data shown in S8 and S9, demonstrate that metal-oxide memristors, preferably in their pristine state, would make simple, accurate passive atmospheric temperature measurement devices. A clear distinction and successful recovery between each temperature can be seen, indicating that this device would make for a suitable temperature sensor in this range. This effect further leads to the possible implementation of temperature as an additional control parameter for a neuromorphic system. To demonstrate this, a system featuring 25 individual metal-oxide memristors acting as input synapses to a single neuron was developed as per the schematic shown in Fig. 3a.

The memristive synapses in this system are temperature controlled, with the raw input to the system being used as feedforward, with a higher total input value resulting in a higher temperature inside the chamber. The weight of each synapse multiplied by its input value are added to an individual accumulator, which produces a spike at the output (Fig. 3b) upon reaching a threshold value, at which point the accumulator is reset. Increasing the temperature of the synapses decreases their corresponding weights, in turn reducing the output spiking rate allowing for a simulation of homeostasis to be performed, as demonstrated in Fig. 3c. This function ignores constant inputs and only provides a change in output upon a change in input, with the polarity of the output change matching the change in input. The output spiking rate (sampled in groups of 25) is held at a constant rate by the balance of temperature and input strength. The change in input strength then results in a temperature change of the synapses, altering their weight to bring the output back to the baseline spiking rate. This concept also holds for a variety of input patterns (see Figs. 3d and e), regardless of the polarity of changes at the input, proving the versatility as well as resilience of the presented approach.

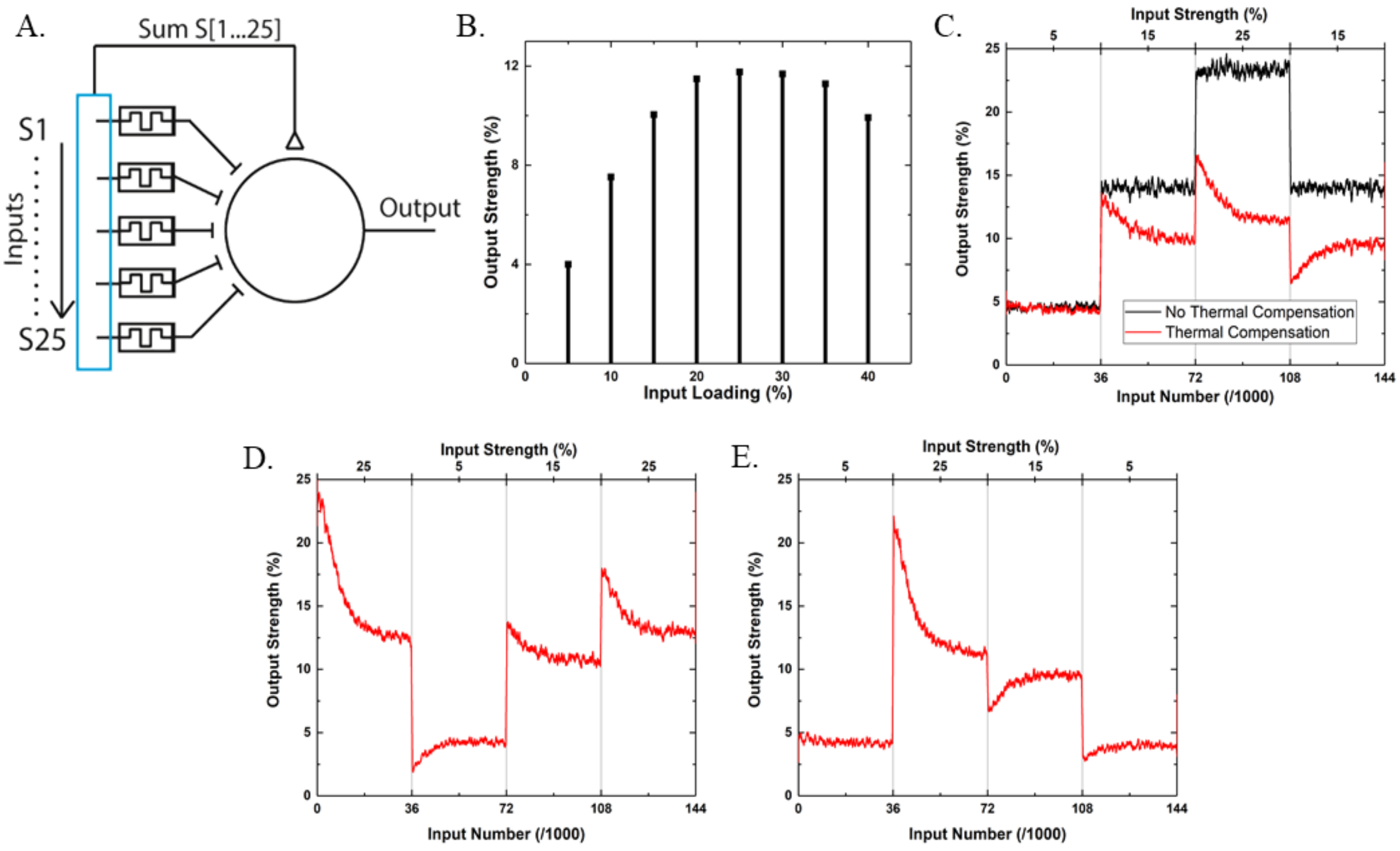

**Fig 3: Emulating homeostatic plasticity via the thermal dependence of memristive weights. a,** Schematic of system comprising 25 input synapses (S1-25) modelled using memristors connected to a single neuron. Feedforward is provided by summing the raw input to the system, controlling the temperature of the memristors. **b**, Baseline output spiking rate of this system with different input

loads, indicating an optimal range for operation of between 15 and 40% input load. **c**, Impact of the thermal compensation applied by the system with a changing input. Changes in the input strength cause the output spiking rate to change rapidly, before moving back to the base rate. **d**, and **e**, Measured results of different exemplar input patterns, showing the system's ability to display both positive and negative changes at the input.

**Conclusion**

In this paper, we have demonstrated that both the static and switching behaviour of metal-oxide memristors are affected by temperature, with sensitivity to switching voltages decreasing with increasing temperature, and device resistivity decreasing under static conditions as a result of Schottky barrier height modulation. This effect has practical applications as demonstrated within the context of a neuromorphic system as an additional method for the control of synaptic weights, allowing for introducing a global control to the system's behaviour. This can either be utilised as the sole method of controlling input weights, or in combination with electrical programming for increasing the functional possibilities. Our study reinforces the necessity for factoring in the thermally induced behavioural changes when designing circuits with such emerging device technologies for applications involving fluctuating temperatures. At the same time, it demonstrates how these behavioural changes can be utilised as a new device approach for sensing temperature.

**Methods:**

The atmospheric temperature around the devices was controlled using the ArC micro-chamber (see supplementary Fig. S1)[20]. Programming the memristors at the resistive states under study was achieved using ArC one's pulsing-based and compliance-free electroforming protocol (see supplementary Fig S3).

The memristor prototypes used throughout this series of experiments consisted of a $TiO_2$ active layer, with a platinum top and gold bottom electrode, both 25 nm thick. This was deposited to an oxidised 6" Si wafer (200 nm thermal $SiO_2$) using a 5 nm titanium layer as adhesion (Figure 1a). The active layer was deposited via reactive magnetron sputtering from a Ti target with 8:35 sccm $O_2$:Ar flow using a Leybold Helios sputtering system. As moisture has been reported to affect sputtered oxide films[21], the micro-chamber was sealed to ensure constant concentration of moisture throughout all experiments.

**Supplementary material:**

**S1: Static Thermal Effects - repeatability**

The protocol applied for experiments shown in Fig. 1 was further applied to a range of devices, covering the following resistive state ranges: pristine (prior to electroforming) state and four post electroforming resistive levels (L1 ~1 MΩ, L2 ~250 kΩ, L3 ~15 kΩ, L4 ~8 kΩ). Discrete devices appeared to adapt variations in temperature in a similar manner and in all cases they were found to converge within the 1 hr time window selected for this experiment (Fig. 1b and Fig. S4 in the supplementary).

In the pristine and L1 post-electroformed resistive levels, we observed a similar and consistent response characterised by an initial, high sensitivity of around 1% resistance change per deg. K (denoted: ΔR/ΔT from initial R at 300 K) with a tendency to saturate at the higher end of the temperature range. At lower resistive levels (L2-L4) we obtain lower sensitivity ΔR/ΔT ∈[0.28%, 0.66%] (Fig. 2a). This feature enables post fabrication tuning of the device temperature response.

Conduction signature analysis performed on the pristine and L1 devices (Fig. 1d) revealed thermionic emission to be the strongest conduction mechanism present, this subsequently explains the presence of both the reduction in resistivity with increasing temperature, while also supporting the non-linearity of the response in these cases.

**S2: Instrumentation and set-up**

This study was supported by developing an environmental Microchamber that allowed evaluating the sensing capabilities of all DUTs. This Microchamber platform allowed for in-situ tuning of the air temperature and relative humidity around the devices. In addition, the setup was able to communicate with our memristors' characterisation platform ArC ONE allowing to perform a comprehensive characterisation including all the figures of merit of interest for memristive technologies. Figure S1, presents the full setup that is also described in detail in[20]. This miniaturised environmental control chamber allows for temperatures between room temperature and 360 K to be applied.

**S3: Understanding the mechanism underneath: initial interpretation**

The influence of temperature on both static and dynamic device responses can be explained by a thermally-activated mechanism dominating their electrical properties. This has been assessed by recording the static IV curves (supplementary Fig. S5) at the same temperature range. For devices operating in their pristine state as well as for the L1, showing the highest sensitivity, detailed analysis revealed signature plots (see supplementary Fig. S6) supporting thermionic emission over the interfacial barriers[22] as the responsible conduction mechanism, described by eq. 1:

$$I = AT^2 \frac{e^{-\phi_b - \alpha\sqrt{v}}}{kT}$$

where $I$ is the current flow through the device, $A$ is a constant that includes the Richardson constant and the effective area of the device, $T$ is the absolute temperature, $\varphi_b$ is the Schottky barrier height at zero applied field, $\alpha$ is the barrier lowering factor, $v$ is the voltage and $k$ is the Boltzmann constant.

As a result, the resistive state obtained by a read-out voltage ($R=v/I$) is the effect of a reversed biased Schottky contact mainly determined by the apparent barrier ($\phi_{App} = \phi - \alpha\sqrt{v}$), eq. 2:

$$R = A'T^{-2}\frac{e^{\phi}_{App}}{kT}$$

Where $A'$ is $1/A$. Consequently, resistive switching arises by the modification of the apparent barrier at the readout voltage due to an electrical stimulus pulse and therefore (eq. 3):

$$dR = A''T^{-3}\frac{e^{\phi}_{App}}{kT}d\phi_{App}$$

Where $A''$ is $A'\cdot k$. Eq. 3 suggests that for a given modulation in the apparent barrier, the induced change at the read-out resistance, decreases with increasing temperature.

Notably, the temperature $T$ refers to the localised temperature at the electronically conducting areas of the Schottky interface (i.e. where the dominant component of current $I$ flows though). This has been indicated as being strongly determined by local heating effects due to high-current densities. As such, corroborating our hypothesis requires a clear indication that substrate temperature affects the temperature at the relevant location within the Schottky contact accordingly.

An additional contributing mechanism might revolve around temperature affecting the conductivity of the electrodes leading to the memristor, and thus introducing a parasitic series resistance. This affects the magnitude of the voltage delivered across the actual device, thus leading to a reduction in the effective $\phi_{App}$. This also leads to a reduction in induced resistance change with increasing temperature. Clarifying the contribution of each mechanism in the ultimate behaviour of the memristive devices with temperature requires dedicated study. The impact of barrier height is unlikely to be observed within static read results due to the low voltage utilised resulting in a very minimal change in barrier height. This will result in the resistance of the device being the dominant reading over the line resistance.

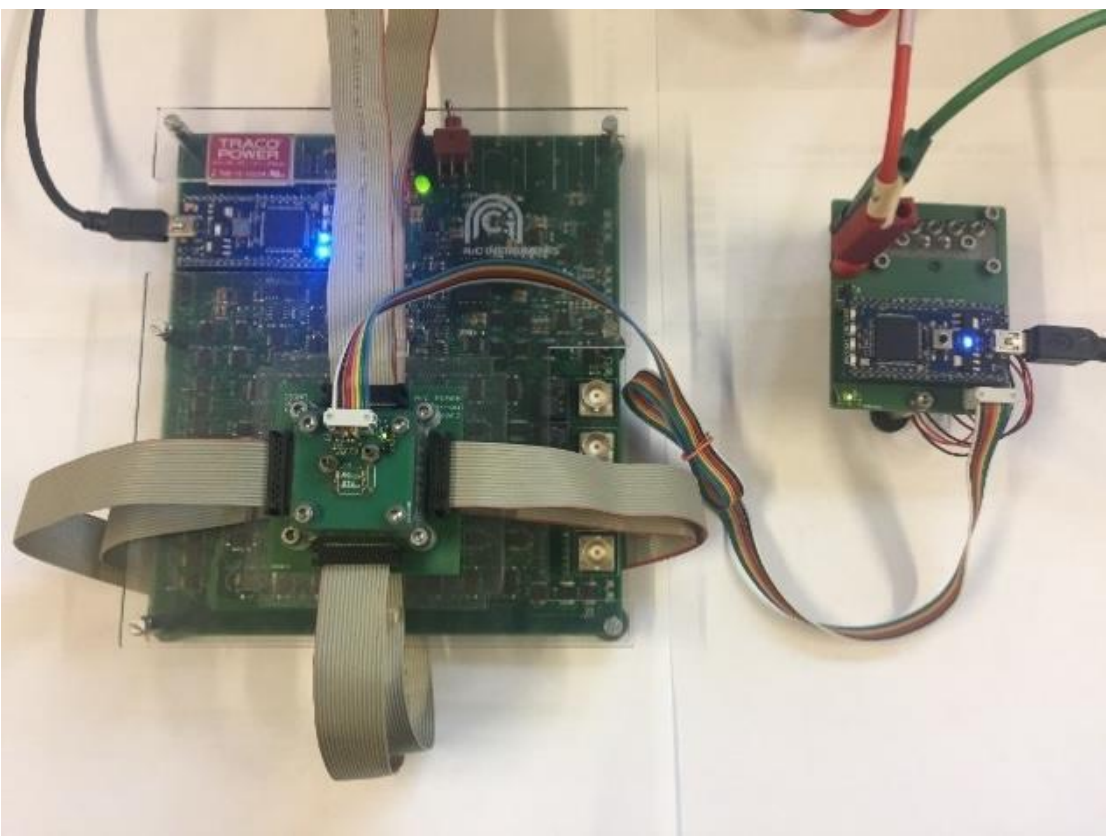

**Figure S1:** The experimental apparatus used to record the data, consisting of an ArC One and an ArC Environmental Microchamber

In order to assess the effect of environmental temperature on the static and switching characteristics of our devices, we established the testing protocols presented in Figs. S2 a and b respectively. These allow for fair comparison between devices operating in different resistive levels.

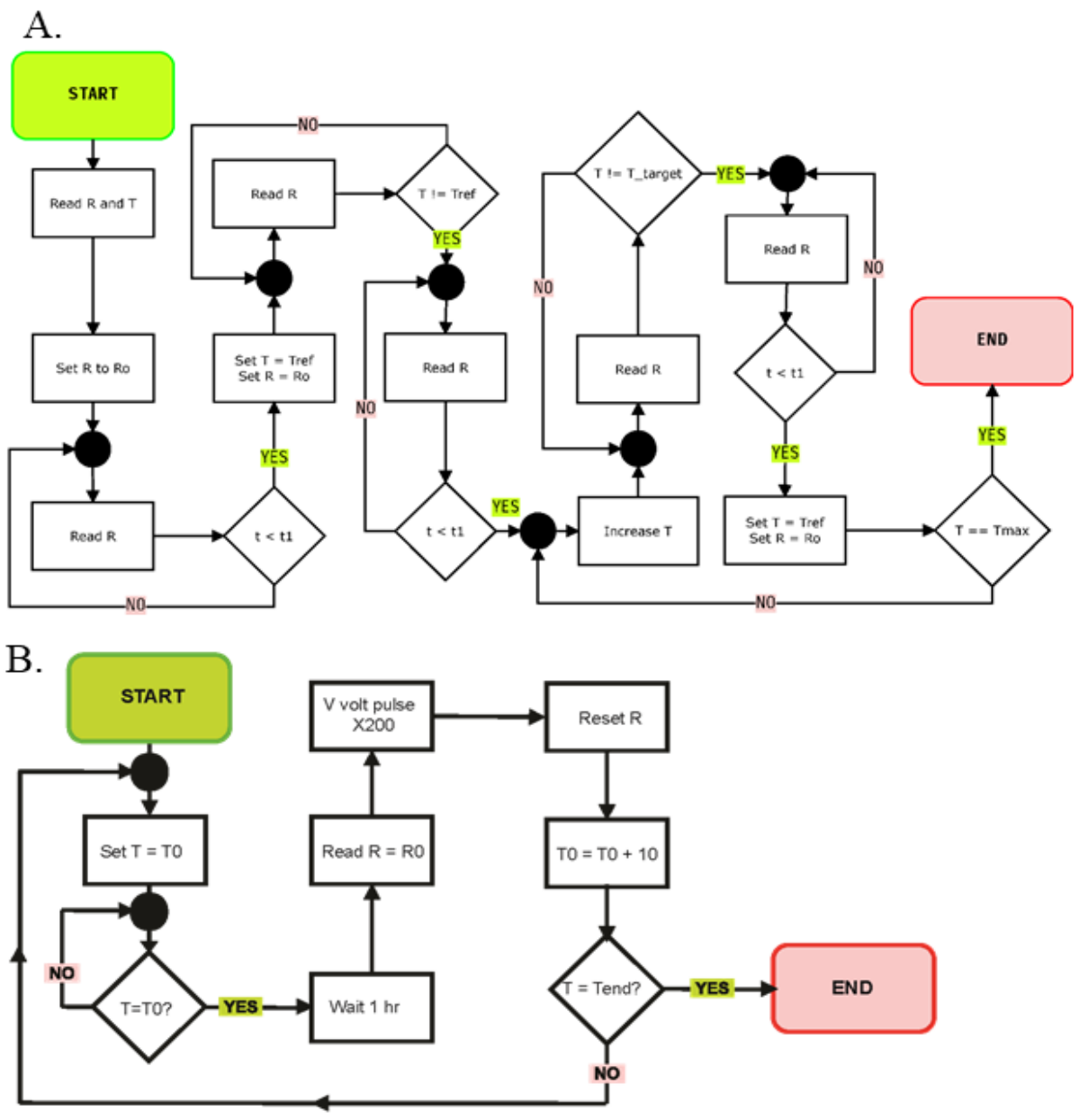

**Figure S2** (a) The protocol for monitoring device resistivity while heating. (b) The protocol used when switching the devices while stepping through temperatures.

The use of Arc-ONE allows for electroforming the devices by a pulsing based and compliance free protocol. This is achieved by applying trains of pulses having progressively increasing amplitude and/or duration across the stacks. The sequence keeps on running until the device reach a pre-defined resistive level. Series resistance could be also potential utilised for further device protection from breakdown. Thanks to the short pulse duration (ranging from 100 ns to 100 ms) this protocol turned out to be gentler that the typically used current-compliance methods and thus enabled the programming of devices on various resistive levels. The algorithm and the representative application of the procedure is depicted in Fig. S3.

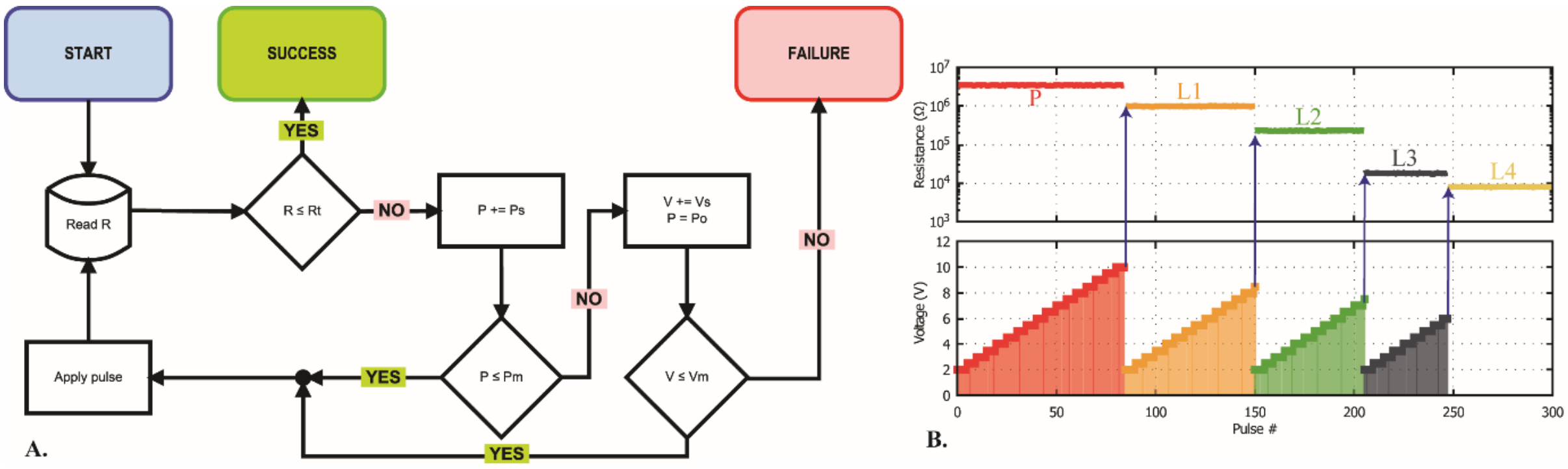

**Figure S3** (a) The algorithm used for forming the devices to their various operating regions, **B** the various resistive levels the devices were formed to and approximate forming voltages.

The employed forming protocol allows for discrete devices to show very similar response on the induced changes in the environmental conditions inside the Microchamber. Fig. S4 shows the thermal convergence for devices at different resistive levels.

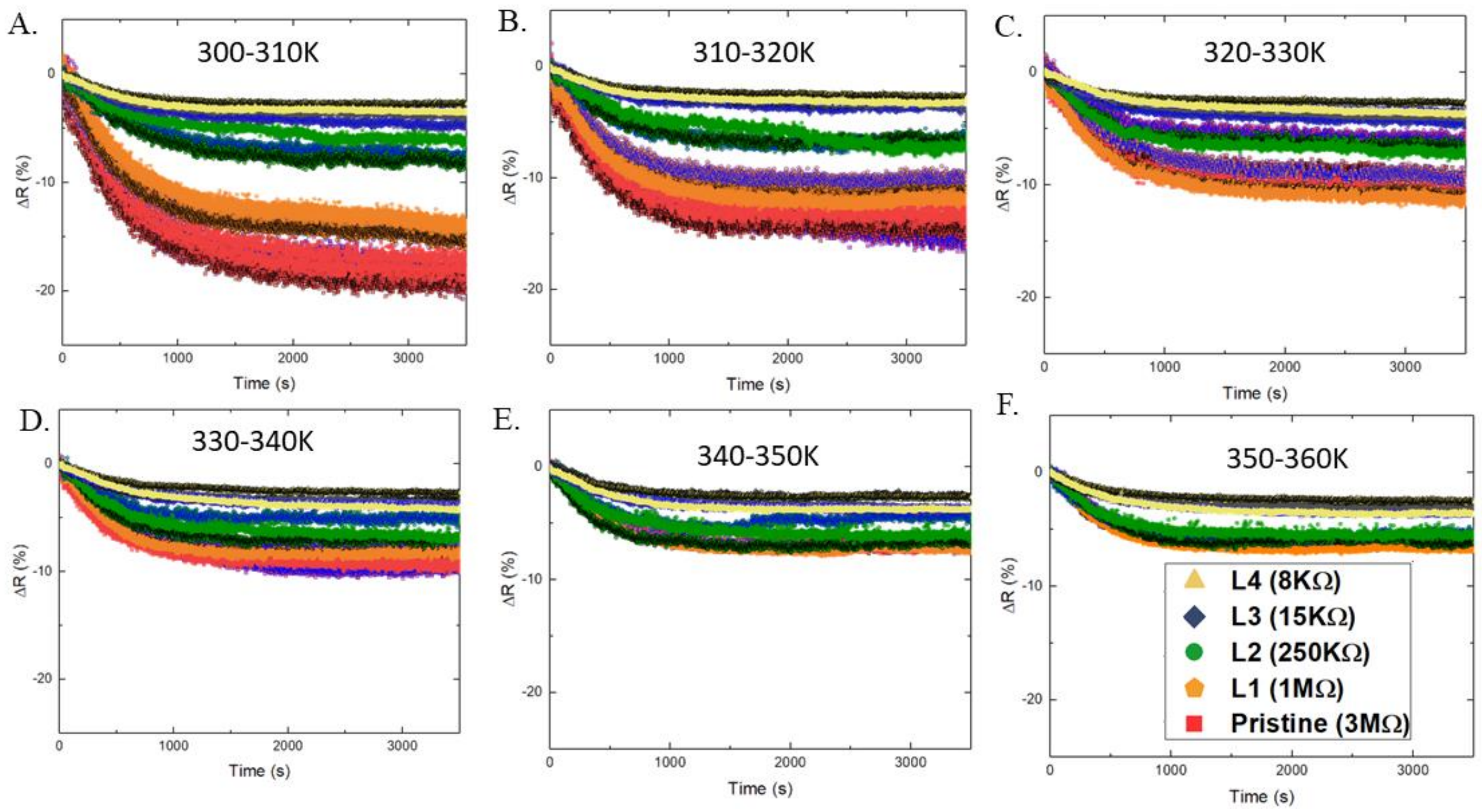

**Figure S4** Thermal convergence for devices in each operating region during heating, showing a reduction in sensitivity to temperature increase with a reduction in resistance **A** From 300 to 310K. **B** From 310 to 320K. **C** From 320 to 330K. **D** From 330 to 340K. **E** From 340 to 350K. **F** From 350 to 360K.

Depending on the resistive level, a specific applied bias value should be reached in order to enable resistive switching effects. Operation of the devices on lower bias levels do not induced changes in their resistive level. We refer to this type of operation as *non-switching* and the corresponding non-switching I-Vs at different temperatures and for the various resistive levels are shown in Fig. S5.

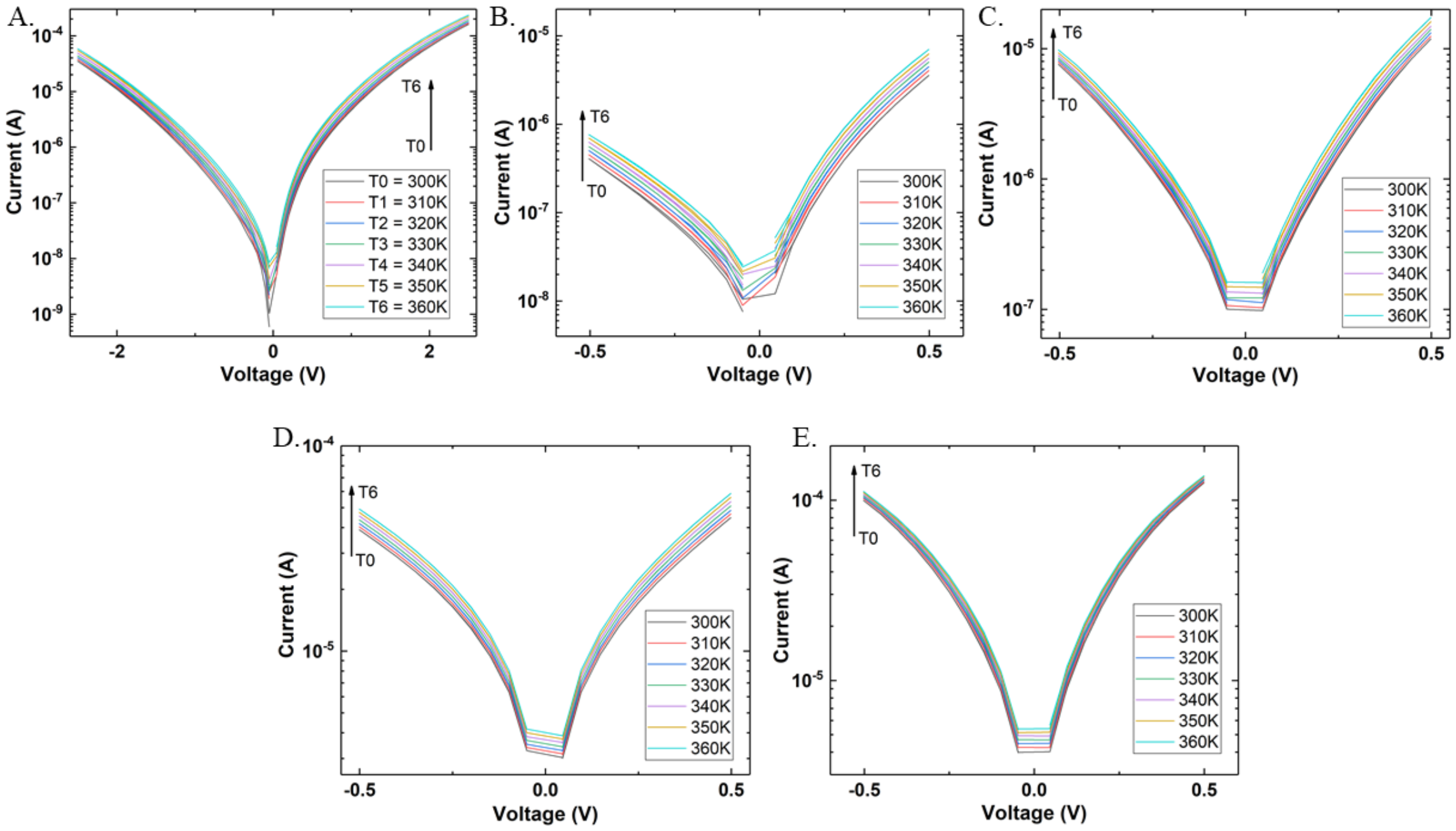

**Figure S5** Non-switching IV curves of $TiO_2$ devices at varying resistive levels. (a) Pristine, (b) Level 1, (c) Level 2, (d) Level 3 and (e) Level 4.

For our $TiO_2$ prototype stacks, the transport in their pristine state as well as in resistive levels corresponds to higher resistances is typically controlled by interface barriers. An asymmetric response in the current voltage characteristics with respect to the applied bias polarity is an initial indication. This is presented in Figs. S5a and S5b. Further evaluation is achieved through the so-called signature plots. This is specific representation of the experimental data in order to verify compliance with the equation that describes the dominant conduction mechanism (Eq. 1).

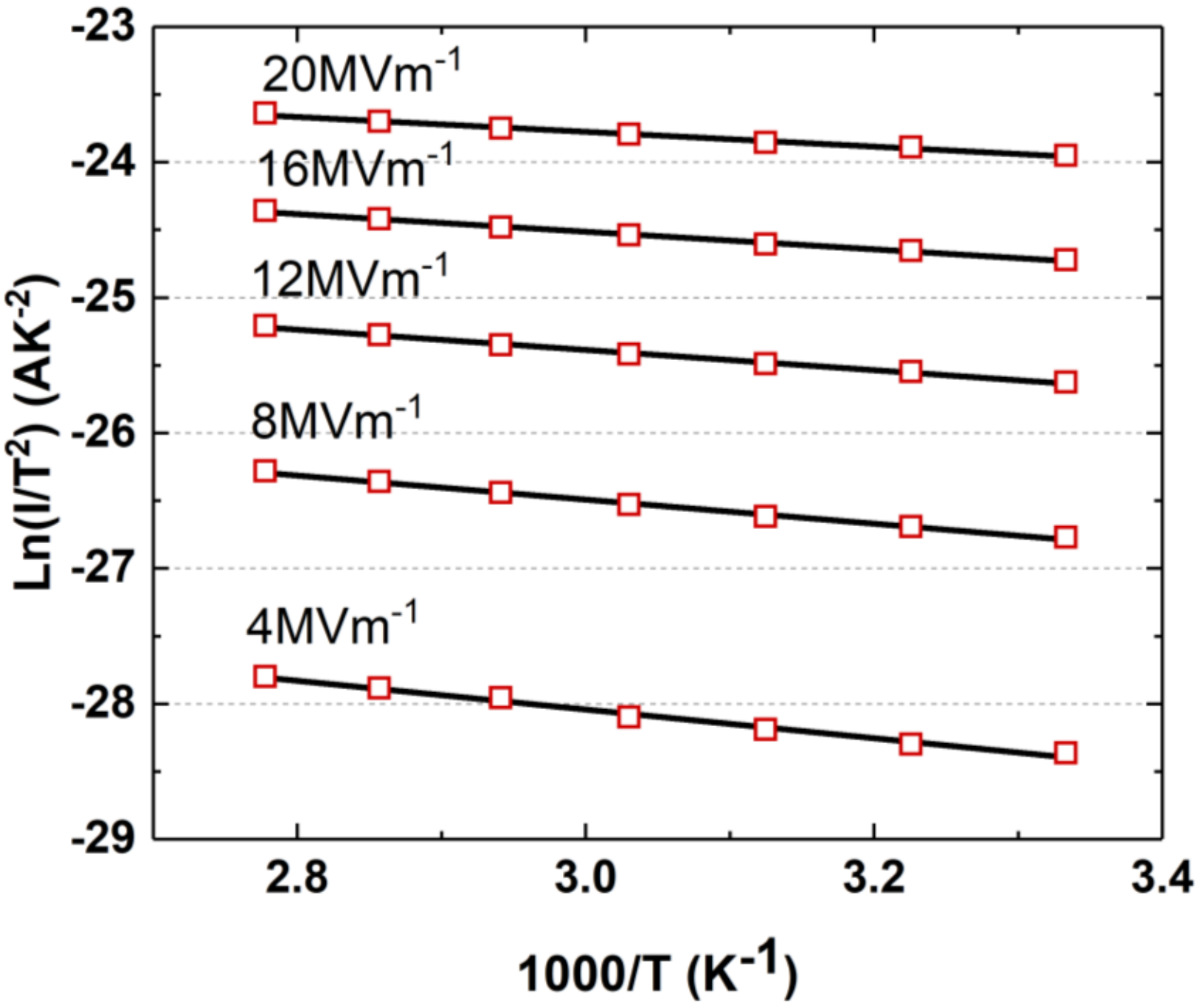

**Figure S6** Signature plot supporting the dominance of thermionic emission in an L1 device.

The neuromorphic application selected was benchmarked to check the linearity of the baseline spiking rate. This shows that the system is capable of operating best between 10 and 40% input loading, with 20-35% being the optimal operating range.

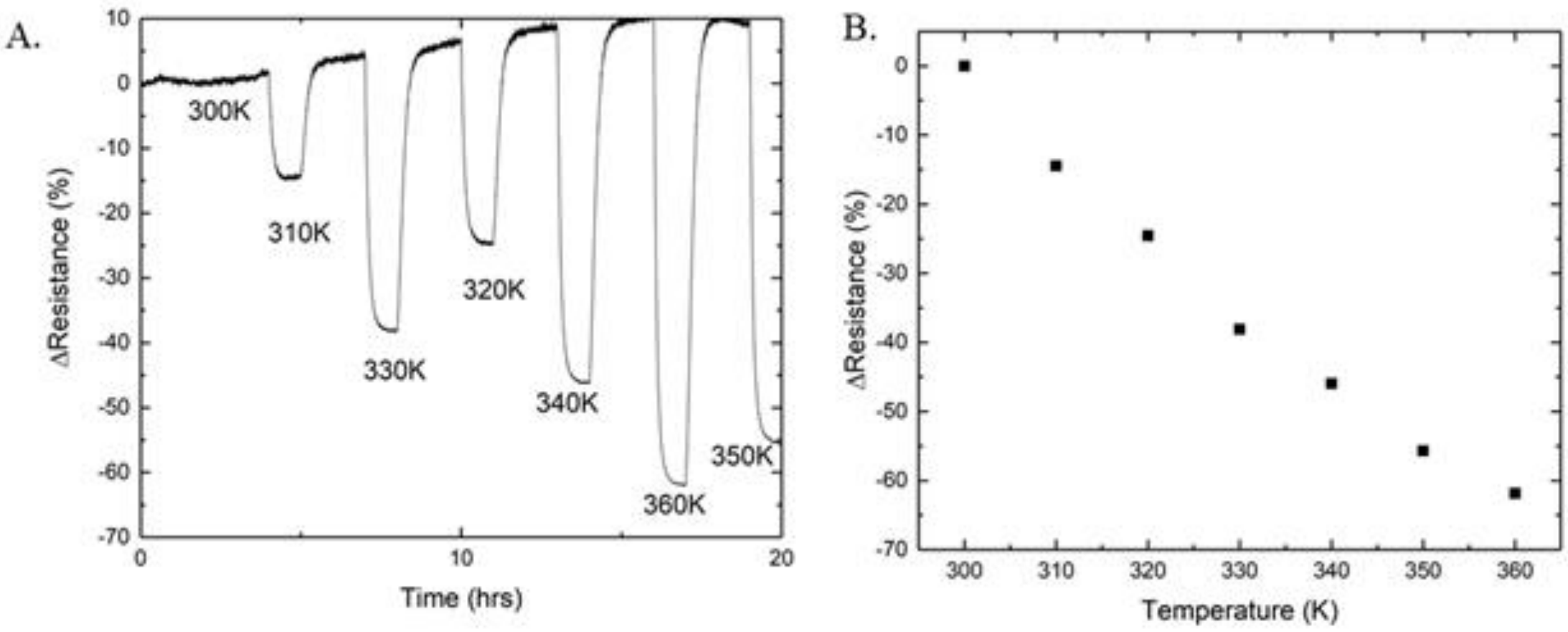

**Figure S7** The response and sensitivity of a pristine device to a randomised heating routine.

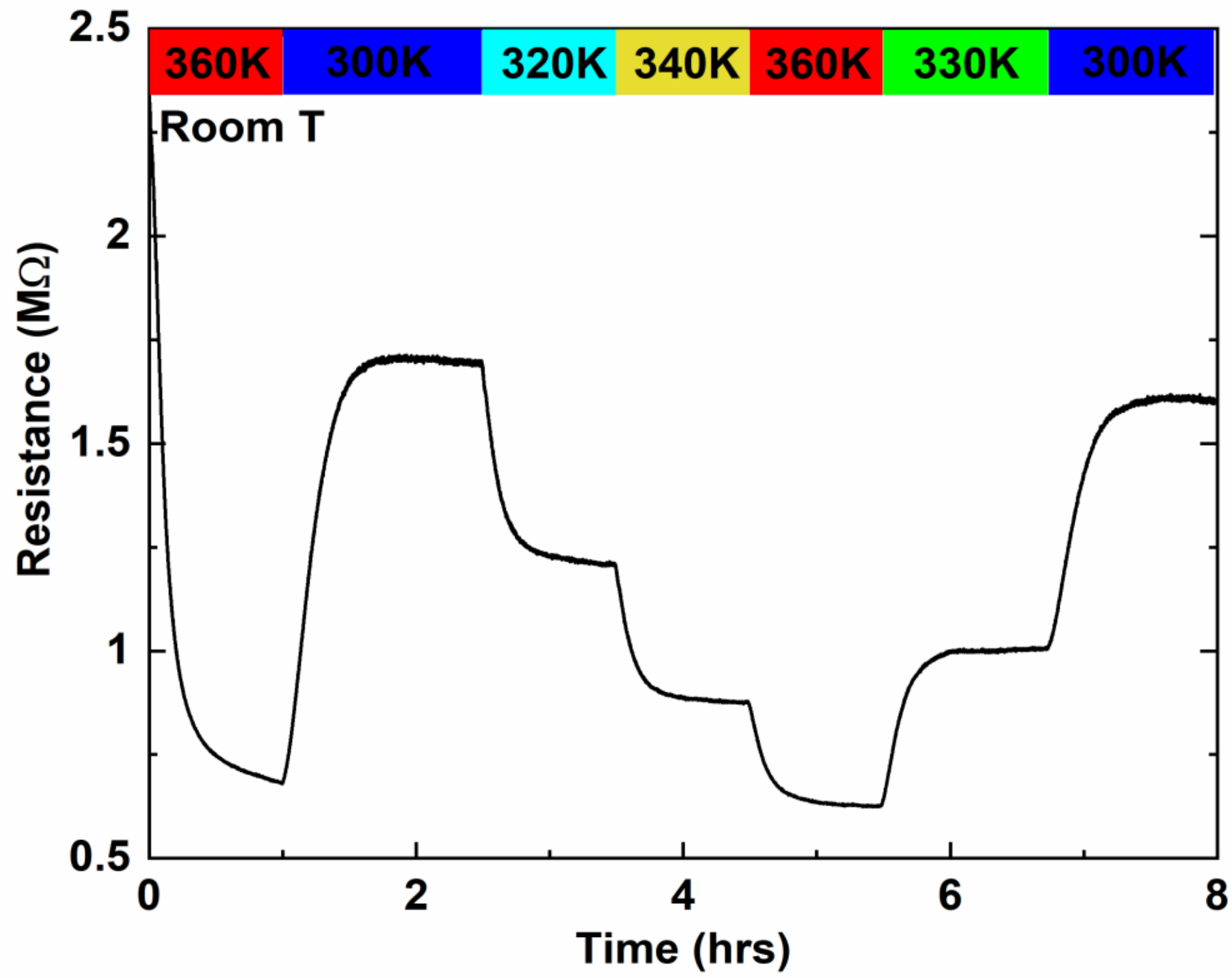

**Figure S8.** Transient response of a metal-oxide memristor's state evaluated under distinct temperatures, showcasing a memristor thermometer. A pristine device used as a passive temperature sensor through a heating cycle schedule with scrambled temperature order showing clear correlation between temperature and measured resistive state. Note relatively small difference (5%) between static resistances measured during the first and second visits to 300 K and 360 K.

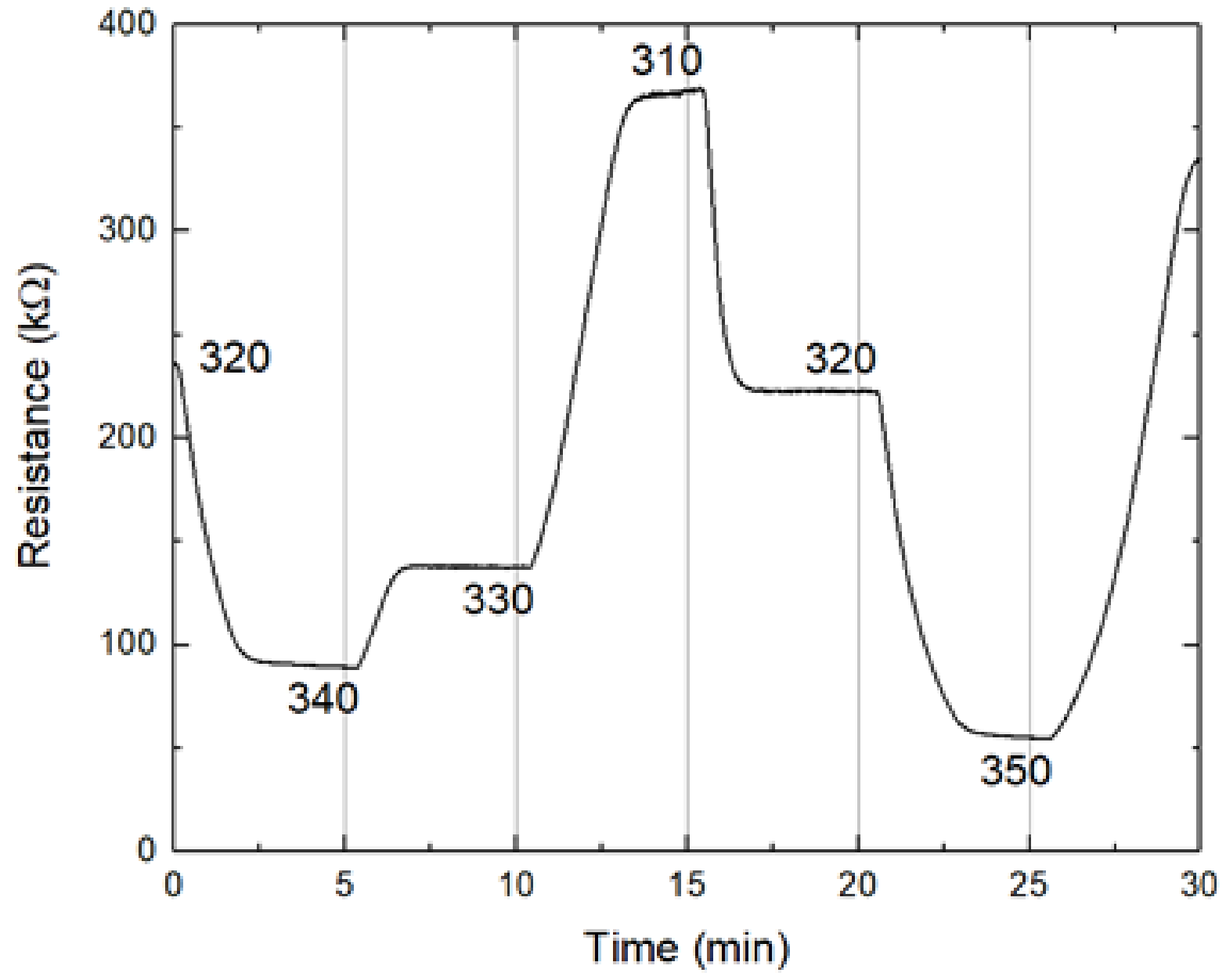

**Figure S9** Transient resistance response of a pristine device on wafer to a randomised heating routine, showing much faster response times to changes in temperature.